\documentclass[12pt]{article}
\usepackage{a4wide}
\usepackage{amssymb}
\usepackage{graphicx}
\begin{document}
{\renewcommand{\thefootnote}{\fnsymbol{footnote}}
\begin{center}
{\LARGE  Critical evaluation of common claims\\[3mm] in loop quantum cosmology}\\
\vspace{1.5em}
Martin Bojowald\footnote{e-mail address: {\tt bojowald@gravity.psu.edu}}
\\
\vspace{0.5em}
Institute for Gravitation and the Cosmos,\\
The Pennsylvania State
University,\\
104 Davey Lab, University Park, PA 16802, USA\\
\vspace{1.5em}
\end{center}
}

\setcounter{footnote}{0}

\begin{abstract}
  A large number of models have been analyzed in loop quantum cosmology, using
  mainly minisuperspace constructions and perturbations. At the same time,
  general physics principles from effective field theory and covariance have
  often been ignored. A consistent introduction of these ingredients requires
  substantial modifications of existing scenarios. As a consequence, none of
  the broader claims made mainly by the Ashtekar school --- such as the
  genericness of bounces with astonishingly semiclassical dynamics, robustness
  with respect to quantization ambiguities, the realization of covariance, and
  the relevance of certain technical results for potential observations ---
  hold up to scrutiny. Several useful lessons for a sustainable version of
  quantum cosmology can be drawn from this outcome.
\end{abstract}

\section{Introduction}
\label{s:Intro}

Loop quantum cosmology is based on potential modifications of space-time
structure such as an underlying discreteness of space. Additional, less
obvious quantum space-time effects may be implied by the resulting
modifications of canonical constraints generating hypersurface deformations in
space-time. A general analysis is therefore expected to be challenging and
rather counter-intuitive. Nevertheless, several very detailed and optimistic
claims have been made about the deep quantum behavior of the theory, in
particular about how the big-bang singularity may be resolved.

These claims paint a picture of quantum space-time that is much simpler than
should have been expected, including for instance a smooth and semiclassical
transition through high curvature: ``Yet, in all cases where the detailed
evolution of {\em quantum} states has been carried out, effective equations
have provided excellent approximations to the full quantum evolution of LQC
[loop quantum cosmology] even in the deep Planck regime, provided the states
are semi-classical initially in the low curvature regime.'' (emphasis in
\cite{Status})\footnote{All quotations from \cite{Status} given in this paper
  refer to the second preprint version.} or ``Indeed, the effective equations
even provide an analytical expression of the maximum density $\rho_{\rm max}$
whose value is in complete agreement with the exact numerical simulations!''
\cite{Status}. And later in the same review, regarding models with positive
spatial curvature, ``For a universe that undergoes a classical recollapse at
$\sim 1{\rm Mpc}$, a state that nearly saturates the uncertainty bound {\em
  initially}, with uncertainties in $\hat{p}_{(\phi)}$ and $\hat{V}|_{\phi}$
spread equally, the relative dispersion in $\hat{V}|_{\phi}$ is still $\sim
10^{-6}$ after some $10^{50}$ cycles.''  (emphasis in \cite{Status}). In the
case of a negative cosmological constant, ``Because the level spacing between
the eigenvalues of $\Theta_{\Lambda}'$ is not exactly periodic, there is a
slight spread in the wave function from one epoch to the next. However, this
dispersion is {\em extremely} small. For a macroscopic universe with
$\Lambda=10^{-120}m_{\rm Pl}^2$, the initially minute dispersion doubles only
after $10^{70}$ cycles!''  (emphasis in \cite{Status}, where
$\Lambda=10^{-120}m_{\rm Pl}^2$ should be $|\Lambda|=10^{-120}m_{\rm Pl}^2$).

For a positive cosmological constant, the evolution operator with respect to a
scalar-field internal time, referred to in \cite{Status} as
$\Theta_{\Lambda}$, is not essentially self-adjoint, unlike the generator
$\Theta_{\Lambda}'$ for a negative cosmological constant.  The statement ``It
is, however, quite surprising that the evolution of such semi-classical states
is largely independent of the self-adjoint extension chosen.'' \cite{Status}
again suggests that quantum effects are severely suppressed for reasons
unknown to \cite{Status}: ``But the precise reason behind the numerically
observed robustness of the quantum evolution is far from being clear and
further exploration may well lead one [sic] an interesting set of results on
sufficient conditions under which inequivalent self-adjoint extensions yield
nearly equivalent evolution of semi-classical states.''

Or, regarding inflationary potentials, a procedure is first set up by ``Now,
we are interested in states which are sharply peaked on a general relativity
trajectory in the regime in which general relativity is an excellent
approximation to LQC. The question is: If we evolve them backward in time
using [a holonomy-modified wave equation] (4.22), do they remain sharply
peaked on the corresponding solution of the effective equations across the
bounce? To answer this question, we need a sufficiently long `time' interval
so that the state can evolve from a density of, say, $\rho=10^{-4}\rho_{\rm
  max}$ where general relativity is a good approximation, to the putative
bounce point and then beyond.'' It is then claimed to imply that ``wave
functions continue to remain sharply peaked on effective trajectories at and
beyond the bounce, independently of the choice of the self-adjoint
extension.''

These statements, all quoted from a single review which, upon close reading,
reveals several troubling features, have often been made within a specific
school of thought within loop quantum cosmology, but they have never been
explained based on general physics principles.  While the review \cite{Status}
is by now rather old, it has been foundational and is still widely
believed to be valid in the very recent literature. Most of the latter did not
question the claims made in \cite{Status} but rather built on them. A detailed
analysis of \cite{Status} therefore remains timely. Other critical viewpoints
have occasionally been presented, such as \cite{ConceptualCosmo,ObsLQC} which
discuss observational questions and the robustness of bounces but do so within
the setting described in \cite{Status}. Others, such as
\cite{VolumeDomain,DressedRevisited}, focus on technical questions of a
specific Hilbert-space representation. Our discussion here will be broader
(and therefore much more damning), pointing out several violations of general
physics principles such as the ubiquity of quantization ambiguities, the
domain of effective field theory, and the condition of general covariance.

In terms of specific constructions, the name ``loop quantum cosmology'' in
current parlance does not refer to a single theory but rather describes a
collection of different frameworks built on the same basic motivation,
originally given in \cite{HomCosmo,IsoCosmo}. The most optimistic, and at the
same time puzzling, claims have been made in what will be referred to here as
the ``Ashtekar school'' because it was initiated by a series of papers
starting with \cite{APS}. These papers and various follow-up studies,
summarized in \cite{Status}, introduced several simplifying assumptions which
initially were not deemed highly restrictive or misleading. One example
is the overly semi-classical nature of evolved states, as already quoted,
which is presented as a surprising result of the general framework but, in
fact, represents only one of several consequences of an erroneous assumption
in the very foundations of the school.

As another example, consider several attempted explanations of singularity
resolution. The abstract of \cite{Status} claims that ``In particular, quantum
geometry creates a brand new repulsive force which is totally negligible at
low space-time curvature but rises rapidly in the Planck regime, overwhelming
the classical gravitational attraction.'' and later ``The key difference
between the WDW [Wheeler--DeWitt] theory and LQC is that, thanks to the
quantum geometry inherited from LQG [loop quantum gravity], LQC has a novel,
built-in repulsive force.'' The review continues by saying that ``For matter
satisfying the usual energy conditions any time a curvature invariant grows to
the Planck scale, quantum geometry effects dilute it, thereby resolving
singularities of general relativity.''  or, more explicitly, ``Thus, the LQC
resolution of the big-bang singularity can evade the original singularity
theorems of general relativity even when matter satisfies {\em all} energy
conditions because Einstein's equations are modified due to quantum gravity
effects.'' (emphasis in \cite{Status}). These claims are puzzling because the
broader singularity theorems of general relativity do not require specific
dynamics but only use properties of Riemannian geometry such as the geodesic
deviation equation. Singularity removal therefore cannot be explained simply
by a new force, which would be a dynamical feature. Or, if there is a new
force that is able to modify the geodesic deviation equation, it would also
have to render space-time geometry non-Riemannian. No such possibility has
been considered by the Ashtekar school.

A deeper analysis reveals that these and other claims made by the Ashtekar
school, percolating through the entire manifesto \cite{Status}, are not only
incorrect but also hint at much deeper problems of the framework. Valuable
lessons can be learned from such an analysis, not only for loop quantum
cosmology itself but also for quantum cosmology in broader terms, as well as
for the full theory of loop quantum gravity or other discrete approaches to
quantum gravity. They also apply to various black-hole models that are based
on replacing the singularity with a bouncing interior within the horizon. Some
observations put together here for a general critique have already been made
before, for instance the importance of infrared renormalization
\cite{Infrared}, the dependence of bounce robustness on a choice of dynamical
representation \cite{NonBouncing}, or the lack of covariance
\cite{TransComm}. In addition to briefly reviewing these observations, we
provide here a comprehensive case that reveals the extensive scope of problems
in loop quantum cosmology, in particular in the Ashtekar school, and we
present an outline of steps that would be required to address these issues.

We briefly summarize the main problems to be discussed: (i) The Ashtekar
school bases its constructions on the erroneous assumption that large comoving
regions may be assumed to be homogeneous in the early universe. As a
justification, an appeal is made to the Belinskii--Khalatnikov--Lifshitz (BKL)
scenario \cite{BKL} without noting that this scenario is valid only
asymptotically close to a singularity and does not provide a generic lower
bound for the size of homogeneous regions. As we will show in
Section~\ref{s:Quanta}, this mistake implies that quantum effects and their
associated quantization ambiguities are severely underestimated. (ii) The
Ashtekar school implicitly makes strong assumptions about the availability of
effective descriptions, thereby rendering any results based on them
non-generic. In particular, the Ashtekar school assumes that a single
effective theory can be used throughout a wide range of vastly different
scales, including the late and the very early universe (and even through a
bounce). In addition, the school assumes that a certain parameter calculated
in a fundamental theory (the so-called ``area gap'') can directly be inserted
in the equations of an effective theory even though no derivation of the
latter from the former theory exists at present. These problems will be
discussed in Section~\ref{s:Eff}. (iii) As shown in Section~\ref{s:Cov}, the
Ashtekar school has made several incorrect claims about the realization of
general covariance in its models.

\section{Quanta of loop quantum cosmology}
\label{s:Quanta}

Key assumptions made by the Ashtekar school downplay the importance of quantum
corrections and quantization ambiguities in loop quantum cosmology. The former
is related to an incorrect interpretation of the averaging volume used to
define homogeneous minisuperspace models; the latter, to an implicit choice of
a dynamical representation which, as it turns out, helps to make bouncing
solutions more prevalent. Quantization ambiguities are also downplayed by an
incorrect understanding of effective theory, which will be the main topic of
the next section. (We note that ambiguities have occasionally been discussed
in the context of the Ashtekar school, for instance the recent revival
\cite{CosmoLor} of effective theories that include higher-order contributions
from the so-called Lorentzian part of the Hamiltonian constraint
\cite{QSDI,IsoCosmo,HubbleSing,AltEffRecollapse}. However, such a
phenomenological analysis takes for granted that there is a semiclassical
bounce and ignores the possibility that quantization ambiguities
may challenge the very existence of a bounce as well as the validity of
certain effective theories.)

\subsection{Loop quantum classicality}

Loop quantum cosmology in all its forms uses connections and triads as basic
variables, given in isotropic and homogeneous models by 
\begin{equation}
 A_a^i=c\delta_a^i \quad,\quad E^a_i=p\delta^a_i
\end{equation}
with time-dependent functions $c(t)$ and $p(t)$. Inserting these fields in the
symplectic term of the full gravitational action,
\begin{equation}
  \frac{1}{8\pi G}\int_{\cal V}  \dot{A}_a^iE^a_i{\rm d}^3x= \frac{3V_0}{8\pi
    G} \dot{c}p 
\end{equation}
integrated over a finite spatial region ${\cal V}$ with coordinate volume
$\int_{\cal V}{\rm d}^3x =V_0$, implies the Poisson bracket
\begin{equation}
 \{c,p\} = \frac{8\pi G}{3V_0}\,.
\end{equation}
(Without loss of generality, we set the Barbero--Immirzi parameter
\cite{AshVarReell,Immirzi} to the value $\gamma=1$.) This bracket depends on
the arbitrary coordinate volume $V_0$ of the averaging region ${\cal V}$. Note
that $V_0$ depends on two conceptually different properties, the choice of the
region ${\cal V}$ as well as spatial coordinates $x$. Assuming that spatial
coordinates have been fixed at an initial stage, we will be concerned only
with the former property, for instance when we restrict ${\cal V}$ to
subregions in a process of infrared renormalization that implies shrinking
$V_0$ in a collapsing universe.

For simplicity, facilitating a direct comparison with \cite{Status}, we will
transform these basic variables to the Hubble parameter ${\cal
  H}=c/\sqrt{|p|}$ and the oriented volume $v=|p|^{3/2}{\rm sgn}p$:
\begin{equation}
 \{{\cal H},v\}= \frac{4\pi G}{V_0}\,.
\end{equation}
The next step introduces modifications motivated by a discrete structure of
space in loop quantum gravity. While loop quantum gravity is a quantum theory
of continuum fields, it leads to discrete spectra of geometrical operators
such as the spatial volume \cite{AreaVol,Vol2,Loll:Vol,Vol}. The corresponding
quantum representation \cite{LoopRep,ALMMT} implies that the Hamiltonian
constraint cannot be quantized directly in its classical form and has to be
modified \cite{RS:Ham,QSDI}. As we will indicate briefly in
Section~\ref{s:Cov}, the brackets of constraints then carry information about
a modified space-time structure, indirectly related to the discreteness of
geometrical spectra. The modifications to be discussed now, initially
motivated by formal properties of a quantum representation, are therefore
indicative of potential dynamical effects implied by discrete space.

In the isotropic context, instead of representing ${\cal H}$ and $v$ in the
standard way of quantum mechanics, only finite shifts in $v$, generated by
${\cal H}$, are represented through holonomy operators
\begin{equation} \label{h}
 \hat{h}_{\mu}= \widehat{\exp(i\mu {\cal H})}\,,
\end{equation}
where $\mu$ is a real parameter with units of length. It is often assumed
that $\mu$ is close to the Planck length, based on common dimensional
arguments, but a precise value remains to be derived from the full theory for
instance through some suitable procedure of coarse graining. The parameter
$\mu$ is not a regulator (but a modifier) because it is not removed in the
quantization procedure used in loop quantum cosmology. As we will discuss in
Section~\ref{s:Eff}, in a proper effective theory $\mu$ may depend on time
through the volume.

The basic commutator then takes the form
\begin{equation} \label{Comm}
 [\hat{h}_{\mu},\hat{v}]= -\frac{4\pi G\hbar \mu}{V_0}  \hat{h}_{\mu}
\end{equation}
and implies a $V_0$-dependent uncertainty relation
\begin{equation} \label{Hv}
 \Delta_{\mu}{\cal H}\Delta v\geq \frac{2\pi G\hbar}{V_0}\,.
\end{equation}
In the absence of an operator directly for $\hat{{\cal H}}$,
we define ${\cal H}$-fluctuations as
\begin{equation}
 \Delta_{\mu}{\cal H}= \frac{\Delta \sin(\mu {\cal H})}{\mu
   \langle\widehat{\cos(\mu  {\cal H})}\rangle}\,.
\end{equation}

In the late universe, large-scale homogeneity implies that the averaging
volume, $V=V_0|v|$, is large. Dividing (\ref{Hv}) by $\langle|\hat{v}|\rangle$,
we therefore have
\begin{equation} \label{HV}
 \Delta_{\mu}{\cal H} \frac{\Delta V}{\langle\hat{V}\rangle}\geq 2\pi
 \frac{\ell_{\rm 
     P}^2}{\langle\hat{V}\rangle} 
\end{equation}
with a tiny minimum
\begin{equation}
 \left(\left(\langle\hat{V}\rangle^{1/3}\Delta_{\mu}{\cal H}\right) \frac{\Delta
     V}{\langle\hat{V}\rangle}\right)_{\rm min}=
 2\pi \frac{\ell_{\rm P}^2}{\langle\hat{V}\rangle^{2/3}} \ll 1
\end{equation}
for the product of dimensionless fluctuations,
$\langle\hat{V}\rangle^{1/3}\Delta_{\mu}{\cal H}$ and $\Delta
V/\langle\hat{V}\rangle$. It is therefore possible to find late-time states
which are very semiclassical, with tiny relative fluctuations of all basic
operators. Such states easily stay sharply peaked for a long time, which is
not surprising because they represent macroscopic objects with huge volumes
$\langle\hat{V}\rangle$. 

However, the same arguments cannot be used at early times, close to the Planck
regime. If we follow the collapse of an initially large-scale homogeneous
universe, while $V$ shrinks, structure forms even within a co-moving volume
${\cal V}$ of constant coordinate size $V_0$. Once inhomogeneity within ${\cal
  V}$ is appreciable, a smaller region should be selected if the collapse
process is still to be tracked using a homogeneous model.  The scale of
homogeneity is therefore progressively reduced as time goes on, without any
lower bound on possible $V_0$ in the classical theory. Nevertheless, there is
still a role played by homogeneous dynamics even close to a spacelike
singularity because the BKL scenario suggests that the generic approach to a
spacelike singularity is locally homogeneous: Close to a spacelike
singularity, time derivatives in Einstein's equation dominate over spatial
derivatives, implying a dynamics approximated by homogeneous
models. Nevertheless, the scale of homogeneity is small, without any general
lower bound. Only the dynamics but not the total homogeneous space of a
Bianchi model can be used as an approximation. In particular, a small
averaging region ${\cal V}$, or a small $V_0$ if spatial coordinates have been
fixed, should be used in order to adjust the dynamics to be of BKL type rather
than of the large-scale homogeneous late-time form. Because the BKL scenario
does not set a lower bound on the scale of homogeneity, it does not justify
any assumption that $V$ should remain much greater than the Planck
volume; it might even be possible to have $V\ll \ell_{\rm P}^3$ well before
Planckian curvature has been reached. (The BKL scenario, being purely
classical, assigns no major role to the Planck length.) The minimal product of
quantum fluctuations implied by (\ref{HV}) is then no longer small, and states
are very quantum.

The Ashtekar school, however, works with the same, large $V_0$ throughout the
entire evolution, thereby suppressing quantum effects in the Planck regime. To
be sure, the averaging volume $V$, given by a constant $V_0$ times the scale
factor cubed, does shrink in a collapsing model.  But because inhomogeneity
generically builds up in any collapsing co-moving volume, the region ${\cal
  V}$, and thus $V_0$, should be shrunk at the same time in order to maintain
the approximation by a homogeneous model.  This reduction of the scale of
homogeneity does not follow from solutions of minisuperspace equations of
motion, but rather from an implementation of infared renormalization in an
effective theory that takes into account limitations of strict minisuperspace
truncations.

There is a subtle mistake in statements such as ``Indeed, according to the BKL
conjecture, the behavior of the gravitational field as one approaches generic
space-like singularities can be largely understood using homogeneous but
anisotropic models. This makes the question of singularity resolution in those
models conceptually important.''  \cite{Status}. It is not the full space of a
homogeneous model that is relevant here but only a tiny region in this
space. Classically, there is no significant difference between these two
interpretations, but the distinction is crucial in quantum cosmology because
uncertainty relations such as (\ref{HV}) depend on the volume. Statements such
as ``For example, for a [closed] universe which grows to a maximum volume of
$1Gpc^3$, the volume at the bounce is approximately $10^{117}\ell_{\rm Pl}^3$!
That the bounce occurs at such a large volume may seem surprising at
first. But what matters is curvature and density and these are {\em always} of
Planck scale at the bounce.'' (emphasis in \cite{Status}) therefore misplace
the wonderment. What is surprising is not the possibility of having a
Planckian density at large volume, but rather the unmentioned assumption that
a generic universe at Planckian density may still be described by a completely
homogeneous space as large as $10^{117}\ell_{\rm Pl}^3$.

Sometimes, the Ashtekar school raises its own puzzling questions in this
context, such as \cite{Status} ``How do quantum gravity corrections manage to
be dominant near the singularity in spite of the fact that the classical
action is large? As we will see, the origin of this phenomenon lies in quantum
geometry.'' and ``For, in the path integral formulation quantum effects
usually become important when the action is small, comparable to the Planck's
constant $\hbar$, while the Einstein--Hilbert action along classical
trajectories that originate in the big-bang is generically very large.'' In
fact, it is wrong to assume that the action of a homogeneous model is
large near the singularity because one should work with microscopic, not
macroscopic ${\cal V}$. The quantum-geometry origin alluded to in the quote is
simply a modification of the classical dynamics, which is unrelated to whether
the action is large or small.

Although parameter estimates given in \cite{Status} rarely refer to the value
of $V_0$ (which is coordinate dependent), they imply large values by the
numbers assigned to the $V_0$-dependent momentum $p_{\phi}=V\dot{\phi}/N$ of a
free massless scalar $\phi$ used for deparameterization: ``The value of the
supremum $\langle\hat{\rho}_{\phi}\rangle$ is directly determined by the area
gap and is in excellent agreement with the earlier studies based on numerical
evolution of semi-classical states. As an example, for semi-classical states
peaked at late times in a macroscopic universe with
$\langle\hat{p}_{(\phi)}\rangle=5000\hbar$, the density at the bound [sic]
already agrees with $\rho_{\rm sup}$ to 1 part in $10^4$. In the k=1 case, for
the universe to reach large macroscopic sizes,
$\langle\hat{p}_{(\phi)}\rangle$, has to be far larger. If we use those values
here, then the density at the bounce and $\rho_{\rm sup}$ would be
indistinguishable.'' Again, in the context of a negative cosmological constant
$\Lambda$, ``Let us consider Schr\"odinger states at a late time which are
semi-classical, peaked at a point on a dynamical trajectory with a macroscopic
value of $p_{(\phi)}$.'' such that ``For macroscopic $p_{(\phi)}$, the
agreement between general relativity and LQC is excellent when $\rho_{\rm
  tot}= \rho+\rho_{\Lambda}\ll \rho_{\rm max}$: departures are significant
only in Planck regimes. In particular, the wave packet faithfully follows the
classical trajectory near the recollapse. Thus, again, LQC successfully meets
the ultraviolet and infrared challenges discussed in section I.'' These
statements show that the same, macroscopic averaging region is used throughout
the entire evolution from low curvature (where the universe is indeed
large-scale homogeneous to a good approximation) all the way to Planckian
curvatures where the only justification for using homogeneous dynamics, given
by the BKL scenario, requires microscopic homogeneous regions.

The erroneous application of macroscopic averaging volumes even at high
curvature is then used to claim robustness of loop quantum cosmology with
respect to quantization ambiguities: ``These examples illustrate that, even
though a priori it may seem that there is considerable freedom in defining the
Hamiltonian constraint, one can introduce well motivated criteria that can
serve as Occam's razor. The two LQC examples we discussed bring out four
important points: i) Internal coherence, a good ultraviolet behavior and the
requirement that quantum dynamics should not lead to large deviations from
general relativity in tame regimes, already constitute powerful constraints;
[\ldots] iv) Although the totality of requirements may seem oppressively
large, they {\em can be} met if one follows a well motivated path that is
conceptually well-grounded.'' (emphasis in \cite{Status}; note that the
reference to Occam's razor in the context of quantization ambiguities is
misplaced because its proper use refers to the number of assumptions invoked
for a specific explanation, while quantization ambiguities do not constitute
independent assumptions).  A closely related problem will be discussed in more
detail in the next section: the claim that a single effective theory should be
used through a wide range of curvature or energy scales.

Also the issue of cosmic forgetfulness \cite{BeforeBB,Harmonic} (see also
\cite{GFTGen} in the context of group-field cosmology
\cite{GFTCosmo,GFTCosmo2,GFTCosmo3,GFTLattice}) is misinterpreted by the
Ashtekar school as a consequence of several misconceptions, in particular
about the role of the averaging region. The authors of \cite{Status} quote a
result from \cite{LoopScattering}:
\begin{equation} \label{Ineq}
 \mbox{``}|\sigma_+-\sigma_-|\leq 2\sigma_*
\end{equation}
where
\begin{equation}
 \sigma_{\pm}= \langle \Delta\ln
 \frac{\hat{V}|_{\phi}}{2\pi\gamma\lambda\ell_{\rm Pl}^2} \rangle_{\pm},
 \quad\quad
 \sigma_*=\langle\Delta\ln(\frac{\hat{p}_{(\phi)}}{\sqrt{G}\hbar})\rangle
\mbox{''} 
\end{equation}
and $\lambda^2= 4\sqrt{3}\pi\gamma\ell_{\rm Pl}^2$.  The inequality is
supposed to restrict the difference of relative volume fluctuations,
$\sigma_{\pm}$, at times well before ($-$) and well after ($+$) the bounce. As
usual, the authors of \cite{Status} ``begin with a semi-classical state in the
distant past well before the bounce'' where a large averaging region can be
used. Then, ``To make the discussion more concrete let us suppose that, when
the hypothetical universe under consideration has a radius equal to the
observable radius of our own universe at the CMB time, it has the same density
as our universe then had. For such a universe $\langle\hat{p}_{(\phi)}\rangle
\approx 10^{126}$ in Planck units. Thus, the coefficient on the right side of
(3.30) [our (\ref{Ineq})] is $\sim 10^{-124}$!'' However, $p_{\phi}$ depends
on $V_0$, and therefore gets successively smaller through infrared
renormalization as the bounce is approached (notwithstanding the fact that
$p_{\phi}$ is a constant of motion in a free-scalar model based on a fixed
$V_0$). The estimate given in \cite{Status} therefore is not applicable in a
discussion of relationships between pre and post bounce values of volume
fluctuations. (The right-hand side of (\ref{Ineq}) is determined by quantum
fluctuations, which generically depend on $V_0$; see (\ref{Hv}).) For a
further discussion of cosmic forgetfulness, see \cite{Casimir}.

In fact, an argument is often used to suggest that a limit $V_0\to\infty$ must
be taken, removing any non-zero lower bound on quantum fluctuations in
(\ref{Hv}). This limit has been suggested because a classical isotropic model
without spatial curvature can be described by an infinite space ${\mathbb
  R}^3$, in which any finite averaging region ${\cal V}$ is only one small
part. The restriction of space to a finite region is then considered an
infrared regulator, which, as is argued, should be removed after quantization
by taking the limit $V_0\to\infty$. According to \cite{Status}, ``As we saw in
section II, a natural strategy is to introduce an infrared regulator, i.e., a
cell ${\cal C}$, and restrict all integrations to it. But we have the
rescaling freedom ${\cal C}\to\beta^3{\cal C}$ where $\beta$ is a positive
real number. How do various structures react to this change? At the classical
level, we found in section II~A that, although the symplectic structure and
the Hamiltonian transform non-trivially, physics ---the equations of motion
for geometry and matter fields--- are all invariant under the rescaling. What
about the quantum theory?'' A regulator implemented in the classical theory
does {\em not} leave physics, that is, the classical solution space,
invariant; this is why a regulator ultimately has to be removed. Sometimes, a
regulator leaves equations of motion invariant, as indicated in the quote, but
then restricts the solution space by imposing boundary conditions, as when a
field is put in a box as a proper infrared regulator. However, choosing a
finite $V_0$ in a homogeneous model based on an infinite space ${\mathbb R}^3$
does not impose any boundary conditions on homogeneous solutions. Changing the
value of $V_0$ does not in any way affect the solution space of a classical
homogeneous model. (The Hamiltonian should of course change under this
transformation because its matter part determines the energy contained in a
region of size $V_0$, which does depend on $V_0$.)  Varying $V_0$ is therefore
a classical symmetry, not the choice of a regulator. Such a symmetry may or
may not be broken by quantum effects.

Turning first to the so-called $\mu_o$-scheme, \cite{Status} continues by
saying ``However, a closer examination shows that this dynamics has several
inadmissible features. First, the energy density at which bounce occurs scales
with the change ${\cal C}\to\beta^3{\cal C}$ in the size of the fiducial
cell. [\ldots] But the density at the bounce should have a direct physical
meaning that should not depend on the size of the cell. Moreover, if we were
to remove the infrared regulator in this final result, i.e., take the limit in
which ${\cal C}$ occupies the full ${\mathbb R}^3$, we would find $\rho_{\rm
  max}^{(\mu_o)}\to0$!'' Mistaking scaling independence after quantization as
a ``theoretical coherence criterion'', \cite{Status} observes that it can be
bypassed ``by restricting oneself to the spatially compact ${\mathbb T}^3$
topology where one does not need a fiducial cell at all.''  Here,
\cite{Status} erroneously suggests that homogeneity of the entire space should
be used in a study of near-singular dynamics. The final conclusion, ``Suppose
that, when its radius equals the observable radius of our own universe at the
CMB time, it has the same density as our universe then had. For such a
universe $p_{(\phi)}\approx 10^{126}$ in Planck units so the density at the
bounce would be $\rho_{\rm max}^{(\mu_o)}\approx 10^{-32} {\rm gm}/{\rm
  cm}^3$!''  \cite{Status}, then makes the mistake of assuming a full
homogeneous space as well as a constant, macroscopic ${\cal V}$ throughout a
wide energy or curvature range. An extrapolation from CMB scales to the Planck
scale within a single homogeneous effective description is illegitimate.

These arguments are flawed for several reasons. In particular, restricting
homogeneous space, be it finite or infinite, to a region ${\cal V}$ does not
constitute regularization because it modifies neither equations of motion nor
the solution space of the theory. Classically, we can work with any finite
$V_0$ and obtain the same predictions for observables. Changing $V_0$ is
therefore a symmetry of the classical theory unrelated to setting a regulator
that would have to be removed. This symmetry is broken by quantum effects when
fluctuations are included, as shown by (\ref{Hv}) which is not covariant with
respect to changing $V_0$. If it is supposedly possible to counter this broken
symmetry by choosing a specific value of $V_0$, such as the volume of a torus,
it must be justified by additional arguments which have not been provided by
the Ashtekar school. Because the symmetry is broken at the quantum level, any
such argument should refer to specific quantum physics.

In fact, $V_0$ does not set an infrared regulator but rather a running
infrared scale that determines the range in which a homogeneous (or
perturbatively inhomogeneous) description can be used. It is then not
surprising that quantum effects computed in a homogeneous model depend on the
scale ($V_0$) on which homogeneity is realized. This running scale
changes from large (truly infrared) values at late times to small (actually
ultraviolet) values at early times, described within a BKL
scenario. Connecting infrared physics to ultraviolet physics in this manner
requires much care, which is completely neglected if one works with a fixed,
large $V_0$ throughout all evolution. How to set up a valid effective field
theory of loop quantum cosmology will be discussed in more detail in the next
section. For now, we will consider the small-$V_0$ regime in more detail in
order to see implications for the possibility of non-bouncing solutions.

\subsection{Loop quantum serendipity}

Proceeding with how a specific version of loop quantum cosmology is set up, we
return to the basic commutator (\ref{Comm}) and see how it may affect the
dynamics. To this end, we should find a dynamical representation in which not
only $\hat{h}_{\mu}$ and $\hat{V}=V_0|\hat{v}|$ operate on wave functions,
respecting (\ref{Comm}), but also a suitable Hamiltonian that encodes the
dynamics of isotropic cosmological models. It turns out that the principles of
loop quantum cosmology can accomodate a large variety of inequivalent
representations, some but not all of which have strictly positive volume on
dynamical solutions in a free scalar model. Such a representation has
serendipitously been chosen by the Ashtekar school, thereby overstating the
prevalence of bouncing solutions.

\subsubsection{Loop quantum cosmology as a discrete affine theory}

Dynamics requires time, which exposes loop quantum cosmology to the problem of
time in quantum gravity \cite{KucharTime,IshamTime,AndersonTime}. The Ashtekar
school does not solve this problem but rather evades it by working with
deparameterization in a fixed choice of internal time, given by a free,
massless scalar $\phi$ with momentum $p_{\phi}$. (Briefly, deparameterization
does not lead to time reparameterization invariance after quantization because
different choices of internal time that may exist in a single model
generically imply inequivalent quantum corrections
\cite{MultChoice,TwoTimes}.)  The Friedmann equation for spatially flat models
then gives rise to the Hamiltonian constraint
\begin{equation}
 C=-\frac{3}{8\pi G} V{\cal H}^2+ \frac{1}{2}\frac{p_{\phi}^2}{V}=0\,.
\end{equation}
Applying deparameterization at the classical or quantum level, we obtain
$\phi$-evolution generated by a Hamiltonian operator
\begin{equation} \label{H}
 \hat{H}_{\phi}= -\hat{p}_{\phi}= \pm\sqrt{\frac{3}{4\pi G}} \widehat{|V{\cal
     H}|}\,. 
\end{equation}

A straightforward quantization of this Hamiltonian would be given in terms of
the dilation generator $\hat{D}=\widehat{V{\cal H}}$, which is well-defined
and self-adjoint in a quantization of the positive real line, $V>0$. It is
therefore a basic operator in affine quantum cosmology
\cite{AffineQG,AffineSmooth,AffineSing,SpectralAffine,MixAffine}, in which the
volume is restricted to positive values. In loop quantum cosmology, both signs
are allowed for the oriented volume $v$, taking into account the orientation
of space. Nevertheless, the dilation generator plays a role because it happens
to be the Hamiltonian (\ref{H}) in a simple deparameterized model.

However, loop quantum cosmology does not provide an operator $\hat{{\cal
    H}}$. A more involved quantization of (\ref{H}) is therefore required,
making use of holonomy operators (\ref{h}) with some non-zero $\mu$ (a
quantization ambiguity). With this modification, the loop-quantum-cosmology
version of (\ref{H}) can be interpreted as a discrete affine quantization of
isotropic cosmology. Instead of basic operators $\hat{V}$ and $\hat{D}$ as in
affine quantum cosmology, it uses basic operators $\hat{v}$ together with
\begin{equation} \label{J}
 \hat{J}_{\mu}= \mu^{-1}\hat{v} \hat{h}_{\mu} \quad,\quad \hat{J}^{\dagger}_{\mu}=
 \mu^{-1} \hat{h}_{-\mu} \hat{v}\,,
\end{equation}
where the factor of $\hat{v}$ demonstrates the kinship to affine quantum
cosmology, while the use of holonomy operators, quantizing periodic functions
of ${\cal H}$, provides discreteness of $v$. (In what follows, we will
suppress the subscript $\mu$ on $\hat{J}$ in order to avoid overloading the
notation.)  Using self-adjoint linear combinations
\begin{equation}
 \hat{J}_+= {\rm Re}\hat{J}=
 \frac{1}{2}\left(\hat{J}+\hat{J}^{\dagger}\right)\quad,\quad \hat{J}_-={\rm
   Im}\hat{J}=\frac{1}{2i}\left(\hat{J}-\hat{J}^{\dagger}\right)
\end{equation}
the basic operators $(\hat{v},\hat{J}_+,\hat{J}_-)$ can be seen to be
generators of the Lie algebra ${\rm sl}(2,{\mathbb R})$ with relations
\cite{BouncePert} 
\begin{equation}
 [\hat{v},\hat{J}_+]= -\frac{4\pi i\ell_{\rm P}^2\mu}{V_0} \hat{J}_-\quad,\quad
 [\hat{v},\hat{J}_-]= 
 \frac{4\pi i\ell_{\rm P}^2\mu}{V_0} \hat{J}_+\quad,\quad
 [\hat{J}_+,\hat{J}_-]=\frac{4\pi i\ell_{\rm P}^2}{\mu V_0}
 \left(\hat{v}-\frac{2\pi \ell_{\rm P}^2\mu}{V_0}\right)\,.
\end{equation}
(The shift of $\hat{v}$ in the last commutator can be eliminated by redefining
the volume.) In simple models of loop quantum cosmology, this Lie algebra is
dynamical because $|\hat{J}_-|$ can be used as a quantization of the
Hamiltonian (\ref{H}) in the deparameterized free-scalar model.

Loop quantum cosmology therefore replaces the Lie algebra of the affine group
of ${\mathbb R}$ by ${\rm sl}(2,{\mathbb R})$. Unlike the Lie algebra of the
affine group, ${\rm sl}(2,{\mathbb R})$ has several different types of
representations, which generically do not preserve the sign of $v$
\cite{Bargmann}. As pointed out in \cite{NonBouncing}, the Ashtekar school has
implicitly selected a specific representation from the positive discrete
series in which no transformations change the sign of $v$ or even map to $v=0$
eigenstates. Bouncing solutions are therefore guaranteed, but only by an
implicit sleight of hand. Had one chosen a representation from the continuous
series (in which, in spite of the name, the $\hat{v}$-spectrum is still
discrete), dynamical operators connecting the two signs of $v$ would have been
included. (Therefore, even if one accepts the modification of the classical
constraint for non-zero $\mu$, bounces are not guaranteed. The modification
implies an ${\rm sl}(2,{\mathbb R})$-structure but does not select a specific
representation.)

\subsubsection{Loop quantum bounceology}

In addition to the choice of a discrete-series representation, the Ashtekar
school has made several assumptions that, with hindsight, make it easier for
solutions to bounce. For this reason, its discussions are about a pre-supposed
bounce and do not constitute an unbiased approach to possible outcomes of
quantum cosmology. Another example is the application of a large averaging
volume even close to the big bang, where large-scale homogeneity is not
generic. Working with a large averaging volume downplays the role of quantum
corrections, as already discussed, because one is then dealing with a
macroscopic object. Next to correction terms from quantum fluctuations,
tunnelling is another quantum effect that is suppressed for macroscopic
objects (large ${\cal V}$) but can easily happen for microscopic objects
(small ${\cal V}$). In particular, even if discreteness of the volume (or a
``brand new repulsive force'' \cite{Status}) may build up a barrier around
$v=0$, it is conceivable that a small homogeneous patch of a quantum universe
can tunnel through it and encounter a singularity, in contrast to a
large-scale homogeneous macroscopic patch.

Bounce statements are often based on upper bounds derived for the matter
density on suitably generic solutions: ``One can show that {\em all} quantum
states (in the dense domain of the volume operator) undergo a quantum bounce
in the sense that the expectation value of the volume operator has a non-zero
lower bound in any state. More importantly, the matter density operator
$\hat{\rho}|_{\phi}$ has a {\em universal upper bound} on ${\cal H}_{\rm phy}$
and, again, it coincides with $\rho_{\rm max}$ [from effective equations].''
(emphasis in \cite{Status}). However, such bounds can be obtained even for
non-bouncing quantum solutions on which the oriented volume $\hat{v}$ attains
zero expectation value and becomes negative, while having non-zero
variance. Non-zero lower bounds for $\langle|\hat{v}|\rangle$, as opposed to
$\langle\hat{v}\rangle$, are easily possible in this situation because the
absolute value is, by definition, never negative, and must then have a
positive expectation value in a spread-out distribution. This fact is obscured
in \cite{Status} (and elsewhere) because the authors repeatedly write
equations such as (3.20) in \cite{Status},
\begin{equation} \label{FvF}
\mbox{``} (F,\hat{\nu}F)_{\rm phy}= \frac{4\lambda}{\sqrt{12\pi G}}
 \int_{-\infty}^{\infty} {\rm d}x |\frac{\partial F(x_+,\phi)}{\partial
   x}|^2\cosh(\sqrt{12\pi G}x) \mbox{''}
\end{equation}
for the physical expectation value of the oriented volume, called $\hat{\nu}$
in \cite{Status}, in a generic state given by the function $F(x_+,\phi)$ where
$x_+=\phi+x$, and $\lambda^2= 4\sqrt{3}\pi\gamma\ell_{\rm Pl}^2$. The
right-hand side is always positive, leading to the mentioned non-zero lower
bound of the volume expectation value. However, it is not the expectation
value of the oriented volume $\hat{\nu}$ written on the left-hand side, which
would be more complicated in this setting; it may certainly be zero or
negative. The volume expectation value on the right-hand side of (\ref{FvF})
and the resulting lower bound of the volume are guaranteed to be non-zero
simply because they are computed for a positive operator with non-zero
fluctuations.

The origin of the lower bound therefore is not some mysterious repulsive
force, as repeatedly claimed in \cite{Status}, but quantum fluctuations which
are present also in the Wheeler--DeWitt or any other quantization. The
difference between the loop and Wheeler--DeWitt quantizations is only that the
dynamics is modified in the former case, such that the minimal volume is
reached at a finite value of $\phi$. In a Wheeler--DeWitt quantization, as
in the classical model, the minimal volume is reached for $\phi\to\pm\infty$, a
limit in which volume fluctuations can, and do, reach zero and no longer
present an obstacle to a zero volume expectation value in the same limit. The
attempted contrast between loop quantum cosmology and Wheeler-DeWitt quantum
cosmology given in \cite{Status} is therefore misleading: ``Thus, in the WDW
theory, a state corresponding to a contracting universe encounters a
big-crunch singularity in the future evolution, and the state corresponding to
an expanding universe evolves to a big-bang singularity in the backward
evolution.'' while in loop quantum cosmology ``It is important to stress that
the bounce occurs for arbitrary states in sLQC at a positive value of
$\langle\hat{V}_{\phi_B}\rangle$ and the resolution of the classical
singularity is generic.'' Even if
$\langle\hat{V}\rangle=V_0\langle|\hat{v}|\rangle$ remains non-zero,
$\langle\hat{v}\rangle$ may be zero, reaching the classical singularity.

Similarly, upper bounds on the energy density can be traced back to quantum
fluctuations, even on solutions on which $\langle\hat{v}\rangle$ reaches
zero. The latter is not prohibited in suitable representations of ${\rm
  sl}(2,{\mathbb R})$, in particular from the continuous series. Fluctuations
are subject to several conditions, including uncertainty relations as well as
reality conditions that follow from the identity
\begin{equation}
 \mu^2 \hat{J}\hat{J}^{\dagger}=\hat{v}^2
\end{equation}
according to the definition (\ref{J}). Taking an expectation value and using
the basic commutators, we obtain
\begin{equation}
 \mu^2\left(\langle\hat{J}_+\rangle^2+\langle\hat{J}_-\rangle^2
+ (\Delta J_+)^2+ (\Delta J_-)^2\right)=
 \left(\langle\hat{v}\rangle- \frac{2\pi\ell_{\rm P}^2\mu}{V_0}\right)^2-
 \frac{4\pi^2\ell_{\rm P}^2\mu^2}{V_0^2}+(\Delta v)^2\,.
\end{equation}
(The $\mu$-dependent constants on the right are subject to quantization
ambiguities.) Identifying $J_-$ with the loop version of the deparameterized
Hamiltonian (\ref{H}), up to a constant factor, which equals $p_{\phi}$, we
can write the reality condition as
\begin{equation} \label{Real}
 \left(V_0\langle\hat{v}\rangle- 2\pi\ell_{\rm P}^2\mu\right)^2- \mu^2V_0^2 
 \langle\hat{J}_+\rangle^2 = \frac{4\pi G}{3} \mu^2
   \left(\langle\hat{p}_{\phi}\rangle^2+ (\Delta p_{\phi})^2\right) +\mu^2
   V_0^2(\Delta J_+)^2-(\Delta V)^2 +4\pi\ell_{\rm P}^2\mu^2\,.
\end{equation}
For large $\langle\hat{p}_{\phi}\rangle$ (in a macroscopic universe), the
first term on the right-hand side dominates. Since it is always positive,
$V_0\langle\hat{v}\rangle$ cannot reach zero, and not even the small but
non-zero $2\pi\ell_{\rm P}^2\mu$. Therefore, zero is avoided by the oriented
volume, not just by the positive volume
$\langle\hat{V}\rangle=V_0\langle|\hat{v}|\rangle$.

However, during collapse, the averaging region and therefore the value of
$\langle\hat{p}_{\phi}\rangle$ must be successively reduced in order to be
consistent with the BKL scenario. There is no general lower bound on
$\langle\hat{p}_{\phi}\rangle$ in this situation, such that the right-hand
side of (\ref{Real}) may be zero or negative, eliminating any restrictions on
possible values of $\langle\hat{v}\rangle$. Fluctuations on the right-hand
side of (\ref{Real}) are restricted by uncertainty relations, but only from
below such that the negative term $-(\Delta V)^2$ may dominate over the other,
positive contributions.

As shown in more detail in \cite{NonBouncing}, uncertainty relations can be
used to derive the inequality
\begin{equation}
 \langle\hat{V}^2\rangle \geq \frac{2\pi G}{3}\mu^2
 \langle\hat{p}_{\phi}^2\rangle\,, 
\end{equation}
again using the relationship between $p_{\phi}$ and $J_-$ in loop models. In a
simple estimate of the scalar energy density, we may replace the classical
expression $\rho_{\phi}= p_{\phi}^2/(2V^2)$ with
$\langle\hat{p}_{\phi}^2\rangle/(2\langle\hat{V}^2\rangle)$. For this
expression, the lower bound on $\langle\hat{V}^2\rangle$ implies an upper
bound
\begin{equation}
 \frac{\langle\hat{p}_{\phi}^2\rangle}{2\langle\hat{V}^2\rangle} \leq
 \frac{3}{4\pi G\mu^2}\,.
\end{equation}
For $\mu$ close to the Planck length, this upper bound is close to the Planck
density. It holds for all solutions, even those on which
$\langle\hat{v}\rangle=0$ is reached. A Planckian upper bound on the energy
density therefore does not imply that the oriented volume avoids the classical
singularity.

The estimate of $\rho_{\phi}$ by
$\langle\hat{p}_{\phi}^2\rangle/(2\langle\hat{V}^2\rangle)$ is subject to
quantum fluctuations, and it depends on how one defines a density
operator. The precise upper bound may therefore change. However, it is
important to note that a Planckian upper bound on the density or a lower bound
on the positive volume can be derived using only equations for quantum
fluctuations, such as the reality condition and uncertainty relations. These
conditions depend on the algebraic structure of the model, encoded in basic
commutators, and are therefore different for Wheeler--DeWitt quantizations in
which such bounds do not exist. There is certainly a difference between loop
quantization and Wheeler--DeWitt quantization regarding the quantum nature of
singularities, but it is implied by the behavior of quantum fluctuations
rather than new repulsive forces in the former case. In particular, the
Planckian value of the upper density bound is only indirectly related to
quantum geometry through the discreteness of $\hat{v}$ in a loop
representation, which modifies the algebraic structure.

\section{Effective field theory}
\label{s:Eff}

Loop quantum cosmology cannot be a fundamental theory because the universe is
not exactly homogeneous. It can therefore be claimed to be valid only as an
effective theory, to within some approximation. While the connection between
loop quantum cosmology and the purportedly fundamental theory of loop quantum
gravity remains loose, the general principles of effective field theory should
be taken into account when one sets up and interprets models of loop quantum
cosmology. Unfortunately, claims made by the Ashtekar school completely ignore
these principles. Instead, they focus on formal aspects such as the
construction of physical Hilbert spaces (``To address these key physical
questions, one needs a physical Hilbert space and a complete family of Dirac
observables at least some of which diverge at the singularity in the classical
theory.''  \cite{Status}) which are important in fundamental theories but play
a less clear-cut role in effective theories.  Working with a fixed physical
Hilbert space prevents the Ashtekar school from implementing a scheme of
infrared renormalization in which one would use macroscopic $V_0$ at low
curvature and microscopic $V_0$ at high curvature in accordance with the BKL
scenario, and it has other drawbacks to which we turn in this section.

In particular, a single effective theory cannot be assumed to be valid
throughout a wide range of regimes, stretching from low curvature to high
curvature and possibly back to low curvature. However, the Ashtekar school
declared this micconception to be one of their founding principles: While
initial statements such as ``Can one construct a framework that cures the
short-distance limitations of classical general relativity near singularities,
while maintaining an agreement with it at large scales?'' \cite{Status} are
innocuous because they leave some freedom in how one interprets ``construct a
framework,'' they quickly evolve into specific claims which use the principle
of a single effective theory to rule out possible quantizations: ``In a
nutshell, while the singularity was resolved in a well-defined sense, the
theory predicted large deviations from general relativity in the low curvature
regime. [...] When this is corrected, the new, improved Hamiltonian constraint
again {\em resolves the singularity and, at the same time, is free from all
  three drawbacks of the $\mu_o$ scheme.}'' (emphasis in \cite{Status}). The
conclusion that ``Already in the spatially homogeneous situations, the
transition from $\mu_o$ to $\bar{\mu}$ scheme taught us that great care is
needed in the construction of the quantum Hamiltonian constraint to ensure
that the resulting theory is satisfactory both in the ultraviolet {\em and}
infrared.'' (emphasis in \cite{Status}) shows an intimate relationship between
a strong claim to rule out ambiguities and the erroneous assumption that a
single effective theory with fixed parameters should be valid throughout a
wide range of scales.

We have already seen that the infrared scale, given by the averaging volume
$V_0$ of minisuperspace models, should be adjusted to the varying homogeneity
scale in a collapsing and possibly reexpanding universe. Along with this
adjustment, parameters of the effective theory, in general, have to be changed
as well, or renormalized.

\subsection{Ineffective theory}

The Ashtekar school not only assumes that a single effective theory with fixed
parameters specifying the discreteness scale remains valid through a wide
range of energy scales, it also uses this erroneous belief as a condition to
eliminate quantization ambiguities. It may, of course, be possible that such
parameter choices can be made in certain situations, and that they fulfill
other conditions set up within the minisuperspace setting. However,
regime-independent choices are not generic in an effective theory, unless they
have been strictly derived from a fundamental theory (which, at present, is
not possible in loop quantum cosmology). Conclusions drawn from the assumption
of using a single effective theory, related mainly to robustness claims in
loop quantum cosmology, are thus invalid. The Ashtekar school thereby
downplays the range of quantization ambiguities.

In addition, the Ashtekar school has made several further mistakes related to
an effective treatment. For instance, it misinterprets certain quantum
corrections that arise from the discreteness scale. We have already seen
holonomy modifications as one consequence of spatial discreteness in loop
quantum gravity. It so happens that these modifications can (sometimes) be
expressed solely in terms of the energy density of matter, related to the
Planck density. In general, however, corrections in an effective theory may
also refer to the discreteness scale of an underlying fundamental
theory. Another example in loop quantum cosmology is given by inverse-triad
corrections \cite{InvScale,InflTest}.  It is difficult to derive the specific
behavior of these corrections while the fundamental behavior remains poorly
understood, but this fact does not allow one to ignore them in an effective
theory.

Using only the length scales available in a simple minisuperspace setting, the
Planck length $\ell_{\rm P}$ together with the averaging volume $V_0$, the
Ashtekar school assumes that any correction related to a length scale, given
by spatial discreteness, must be determined by the dimensionless parameter
$\ell_{\rm P}^3/V_0$. It then concludes that these corrections are either
meaningless (because $V_0$ is not a physical parameter) or zero (because $V_0$
should be sent to infinity, removing an infrared regulator): ``Numerical
simulations show that, if we use states with values of $p_{(\phi)}$ that
correspond to closed universes that can grow to macroscopic size, and are
sharply peaked at a classical trajectory in the weak curvature region, the
bounce occurs at a sufficiently large volume that these inverse scale factor
corrections are completely negligible.'' \cite{Status} in spatially closed
models, while in spatially flat models, ``If one rescales the cell via ${\cal
  C}\to\beta^2{\cal C}$, for the classical function we have $|\nu|^{-1/2}\to
\beta^{-1} |\nu|^{-1/2}$ while the quantum operator has a complicated
rescaling behavior. Consequently, the inverse volume corrections now acquire a
cell dependence and therefore do not have a direct physical meaning [\ldots]
What happens when we remove the infrared-regulator by taking the cell to fill
all of ${\mathbb R}^3$? Thus, the right side of (4.12) [inverse-volume
corrections] goes to 1.'' The misinterpretation of the infrared scale as a
regulator is made explicit by ``when the topology is ${\mathbb R}^3$, while we
can construct intermediate quantum theories tied to a fiducial cell and keep
track of cell dependent, inverse volume corrections at these stages, when the
infrared regulator is removed to obtain the final theory, these corrections
are washed out for states that are semi-classical at late times.'' in
\cite{Status}.  Both interpretations ignore the correct role of $V_0$ as a
running infrared scale. Corrections referring to the spatial discreteness
scale should also be running, but methods to derive the relevant parameters do
not yet exist in loop quantum gravity. This fact constitutes an incompleteness
of loop quantum cosmology at present; it does not mean that inverse-triad
corrections are meaningless or negligible.

Even the supposedly simple holonomy modifications, which in some cases can be
formulated solely in terms of the energy density without reference to $V_0$,
have been implemented incorrectly in effective theories used by the Ashtekar
school. The usual claim is that holonomy modifications work by replacing the
classical ${\cal H}$ in the Friedmann equation or Hamiltonian constraint with
a periodic function such as $\sin(\mu {\cal H})/\mu$ which can be represented
by holonomy operators. There are two relevant choices in this modification,
given by the value of $\mu$ and by the specific function of $\dot{a}$ (or the
canonical connection component $c$) and $a$ in which the replacement is
periodic. (There is also a choice in the specific periodic function, which
need not be a sine. This choice depends on details of the dynamics,
see for instance \cite{IsoCosmo,HubbleSing,AltEffRecollapse,CosmoLor}, rather
than general considerations of effective theory.)

The Ashtekar school fixes the first choice by declaring that $\mu$, which is
related to the discrete step size in $v$ generated by holonomies as shift
operators, should be given by the smallest non-zero area eigenvalue in loop
quantum gravity: ``Quantum geometry of LQG tells us that at each intersection
of any one of the edges with $S_{12}$, the spin network contributes a quantum
of area $\Delta\ell_{\rm Pl}^2$ on this surface, where
$\Delta=4\sqrt{3}\pi\gamma$. For this LQG state to reproduce the LQC state
$\Psi_o(\mu)$ under consideration, \ldots'' and ``The size of the loop, i.e.,
$\bar{\mu}$ was arrived [at] using a semi-heuristic correspondence between LQG
and LQC states. This procedure parachutes the area gap from full LQG into LQC;
in LQC proper there is no area gap.''  \cite{Status} postulate a very direct
relationship between parameters and states in the fundamental and effective
theories. Calling the semi-heuristic procedure by a fancy name (parachuting)
is not sufficient to overcome the lack of any justification for using a
fundamental parameter directly in an effective theory. The same review
cautions that ``Similarly, it is likely that, in the final theory, the correct
correspondence between full LQG and LQC will require us to use not the 'pure'
area gap used here but a more sophisticated coarse grained version thereof,
and that will change the numerical coefficients in front of $\bar{\mu}$ and
the numerical values of various physical quantities such as the maximum
density we report in this review. So, specific numbers used in this review
should not be taken too literally; they only serve to provide reasonable
estimates, help fix parameters in numerical simulations, etc.'' However, no
justification is given that the ``parachuted'' values may justifiably be
considered reliable estimates of properly coarse-grained values. In
particular, in the presence of large dimensionless numbers, such as the number
of individual patches in a discrete state for a given region, it is not clear
that coarse-grained values should be close to fundamental parameters even in
terms of orders of magnitude.

The second choice, using ${\cal H}=c/\sqrt{|p|}$ as opposed to some other
phase-space function as the argument of holonomies, has several a-posteriori
justifications, mainly given by (i) the implicit assumption that a single
effective theory should be valid throughout a wide range of energy scales and
(ii) the desire to have corrections independent of $V_0$: ``Note that the full
set of effective equations has two key properties: i) they are free from
infrared problems because they reduce to the corresponding equations of
general relativity when $\rho\ll\rho_{\rm max}$; and ii) they are all
independent of the initial choice of the fiducial cell. Even though they may
seem simple and obvious, these viability criteria are not met automatically
but require due care in arriving at effective equations.'' \cite{Status}. The
claim that ``The functional dependence of $\bar{\mu}$ on $\mu$ on the other
hand is robust'' \cite{Status} is based on an erroneous interpretation of the
averaging volume: ``Consequently, the quantum Hamiltonian would have acquired
a non-trivial cell dependence and even in the effective theory (discussed in
section IV) physical predictions would have depended on the choice of ${\cal
  C}$.'' And again, ``Reciprocally, if in place of $\bar{\mu}\sim \mu^{-1/2}$
we had used $\bar{\mu}=\mu_o$, a constant, the big-bang would again have been
replaced by a quantum bounce but we would not have recovered general
relativity in the infrared regime. Indeed, in that theory, there are perfectly
good semi-classical states at late times which, even [sic] evolved backwards,
exhibit a quantum bounce at density of water!''

We have already seen that both arguments in favor of using ${\cal
  H}=c/\sqrt{p}$ follow from an erroneous understanding of effective
theory. The justification of the first choice --- the value of $\mu$ --- is
equally wrong, based on the claim that a parameter of a fundamental theory
should directly determine an only conceptually related parameter of an
effective theory. The argument of the modification function $\sin(\mu {\cal
  H})/\mu$ is therefore not fixed at all but, pending a derivation from loop
quantum gravity, remains ambiguous in both the value of $\mu$ and the
phase-space dependence (even if the sine function is accepted).

From its inception, the Ashtekar school has drawn a sharp dividing line
between two quantization schemes, called the $\mu_0$ and
$\bar{\mu}$-schemes. The $\bar{\mu}$-scheme corresponds to the modification
just discussed, given by $\sin(\mu {\cal H})/\mu$ just writing $\bar{\mu}$
instead of $\mu$. The $\mu_0$-scheme has a similar ambiguity parameter
$\mu=\mu_0$ but has periodicity in $c$ rather than ${\cal
  H}=c/\sqrt{|p|}$. However, in the absence of a strict derivation from loop
quantum gravity, there is a large number of choices of phase-space functions,
which moreover may differ in the various regimes in which effective theories
may be valid. In keeping with common practice in cosmology, usually applied to
the equation-of-state parameter $w=P/\rho$ of matter which need not be
constant but may be assumed to be a piecewise-constant succession of different
values for different power-laws of $\rho(a)$, one may parameterize holonomy
modifications as  \cite{InhomLattice}
\begin{equation} \label{hmux}
 h_{\mu,x}(p,c)=\frac{\sin(\mu |p|^x c)}{\mu |p|^x}\,.
\end{equation} 
The new ambiguity parameter $x$ formally plays the role of $w$. Different
constants $x$ give different power laws for the $a$-dependence of the
periodicity, but a single constant should generically be valid only in a
certain regime of an effective theory, determined by a suitable density
range. The two distinct cases highlighted by the Ashtekar school therefore
hide the appearance of a continuum of an ambiguity parameter characterizing
different power-law behaviors.

The Ashtekar school claims that the $\mu_0$-scheme has been ruled out because
it may imply bounces ``at density of water'' \cite{Status}. This claim, again,
is based on the erroneous assumption that a single effective theory should be
valid through a wide range of energy scales. Moreover, it assumes that $\mu_0$
is determined by the smallest non-zero area eigenvalue in loop quantum
gravity, an unjustified assumption because it postulates that a parameter of a
fundamental theory should directly provide the value of a parameter in an
effective theory.

\subsection{A good run of loop quantum cosmology}

How should an effective theory of loop quantum cosmology be set up? Deriving a
strict effective theory from loop quantum gravity (or some other fundamentally
discrete theory of quantum gravity) is challenging. However, several crucial
ingredients are known at least from general considerations of effective theory
applied to the context of quantum cosmology. In particular, the prevalent
application of minisuperspace models in quantum cosmology can be meaningful,
provided one takes into account the correct role played by the averaging
region ${\cal V}$ as an infrared scale (but not an infrared regulator). This
role can be inferred from analogous constructions \cite{MiniSup} in the much
better-understood case of scalar quantum field theory on a flat background
space-time, where the Coleman--Weinberg potential \cite{ColemanWeinberg}
provides crucial insights.

In this setting, the infrared scale ${\cal V}$ determines the dividing line
between an inner region, given by the physical scale of homogeneity, in which
a minisuperspace model or such a model together with perturbative
inhomogeneity may be used, and the surrounding space in which non-perturbative
inhomogeneity (or a full fundamental theory) would have to be applied. The
infrared scale is not constant but changes according to the physical scale of
homogeneity: As inhomogeneity builds up in the inner region, ${\cal V}$ should
be decreased, pushing more and more modes into the surrounding region in which
a full field theory is applied. It is then not surprising, and certainly not
inconsistent, for physical parameters derived in minisuperspace models, such
as critical energy densities at which certain dynamical features happen, to
depend on the infrared scale ${\cal V}$, or on the volume $V_0$ in a fixed set
of spatial coordinates. This scale, after all, determines how many modes of
the full theory have been approximated by a minisuperspace model. The
collective quantum corrections implied by these modes (such as the infrared
contribution to the Coleman--Weinberg potential in well-understood situations)
then naturally depend on the infrared scale. Since the minisuperspace
dynamics, or an early-universe model following the BKL scenario, reacts only
to these quantum corrections but ignores corrections from strong inhomogeneity
on larger scales, its dynamics depends on the infrared scale. Statements such
as ``But the density at the bounce should have a direct physical meaning that
should not depend on the size of the cell.''  \cite{Status} are therefore
inconsistent with a proper interpretation of the cell as a running infrared
scale.

A detailed formulation requires a fully developed numerical (loop) quantum
cosmology which not only evaluates the straightforward minisuperspace
equations but at the same time tracks how inhomogeneity builds up through
numerical relativity. The latter then tells one whenever the scale of ${\cal
  V}$ has to be adjusted in order to maintain a valid interior minisuperspace
model. Parameters of this model run along with ${\cal V}$, in particular the
values of $\mu$ and $x$ introduced in holonomy modifications (\ref{hmux}) of
loop quantum cosmology. How they run is a difficult question, which will
eventually have to be answered by a derivation of effective equations from
full loop quantum gravity, based on the form and dynamical behavior of
discrete states. (Only limited information is currently available on this
question in loop quantum gravity, for instance from studies of renormalization
in this setting
\cite{BackgroundRenorm,TensorRenorm,SpinFoamPhase,CoarseGrainVertex,HypercubeRenorm,SpinFoamRenormSymm}.)
Even though the construction of such a complete effective theory remains out
of reach, the sketch given here demonstrates that a dependence of physical
parameters on the infrared scale is only a problem of pure minisuperspace
models in which the infrared scale is unknown. However, the scale would be
determined by the validity conditions for homogeneous approximations in small
regions of an inhomogeneous simulation of quantum cosmology.

In early-universe cosmology, making use of the BKL scenario, the infrared
scale will ultimately have to be moved into the ultraviolet, at which point
the minisuperspace plus perturbative treatment is likely to lose its
validity. Nevertheless, the approach to high curvature can be studied within
this effective theory of quantum cosmology.

Attempts \cite{Recover,Imp} have been made to derive parameters of effective
equations, such as $\mu$ and $x$ in (\ref{hmux}) (or $\bar{\mu}$) from
spin-network states. However, current methods are insufficient for two
reasons. First, they generally make use of regular lattice states, which
implies an additional selection of states compared with full
spin-networks. Secondly, they often work in a deparameterized setting, using
the same free massless scalar $\phi$ applied in minisuperspace models. There
is therefore an implicit choice of spatial slices, given by constant $\phi$,
fixing the space-time gauge. Such a choice is valid only if the full theory of
loop quantum gravity is slicing independent and covariant, which brings us to
our last topic.

\section{Covariance}
\label{s:Cov}

The Ashtekar school often assumes that Riemannian space-time is realized in
modified models, even if no attempt has been made to derive the correct
space-time structure. (Recall also that this assumption is inconsistent with
the claim that singularities are resolved even if positive-energy conditions
are respected; see Section~\ref{s:Intro}.) For instance, ``one can show that
this integral [$\int_0^{\tau} d\tau R_{ab}u^au^b$] is finite in the isotropic
and homogeneous LQC, {\em irrespective of the choice of equations of state}
including the ones which lead to a divergence in the Ricci scalar. Thus, the
events where space-time curvature blows up in LQC are harmless weak
singularities.'' (emphasis in \cite{Status}) refers to the Ricci tensor in a
situation which is clearly modified compared with general relativity, such
that the applicability of this classical object is unclear. The only argument
in favor of classical space-time structures is erroneous, using
Palatini-$f(R)$ models as supposed analog actions: ``However, if one
generalizes to theories where the metric and the connection are regarded as
independent, one {\em can} construct a covariant effective action that
reproduces the effective LQC dynamics.''  (emphasis in \cite{Status}). This
claim is wrong: While it is possible to model some modifications in isotropic
models of loop quantum cosmology using Palatini-$f(R)$ actions
\cite{ActionRhoSquared}, any such theory applied to vacuum models, such as
empty Bianchi models or the Schwarzschild interior described by a
Kantowski--Sachs model, implies corrections to the Einstein--Hilbert action
that merely amount to a modified cosmological constant \cite{PalatinifR}. It
is therefore impossible for Palatini-$f(R)$ theories to describe the whole set
of homogeneous models in loop quantum cosmology. Therefore, they cannot be
used as a justification of covariance. (Notice that \cite{PalatinifR} was
published several years before \cite{Status} and \cite{ActionRhoSquared}. The
key lesson from \cite{PalatinifR} in this context is implicitly contained
already in the original paper \cite{Buchdahl} that introduced $f(R)$-gravity
in 1970.)

Describing perturbative inhomogeneity, \cite{Status} mentions that some
approaches use ``suitable gauge fixing to make quantization tractable'' which
makes it impossible to obtain a full understanding of covariance, an off-shell
property that is obscured by gauge-fixing. In the same context, \cite{Status}
praises ``The resulting Hamiltonian theories'' because they ``exhibit a clean
separation between homogeneous and inhomogeneous modes'' even though these
modes play crucially inter-related roles in the hypersurface-deformation
brackets \cite{DiracHamGR,Regained} that govern covariance in the Hamiltonian
formulation. (A background transformation of time and a small inhomogeneous
normal deformation of a spatial slice commute only up to a spatial
diffeomorphism on the slice, forming a semidirect product of Lie algebroids;
see \cite{NonCovDressed} for further discussions.) These modes should {\em
  not} be separated in a covariant model.

In \cite{Transfig}, an attempt was made to extend the methods and ideas
reviewed in \cite{Status} to black-hole models, using the well-known feature
of the Schwarzschild or Kruskal space-time being spatially homogeneous inside
the horizon. These constructions suffer from the same problem discussed here
in cosmological models, assuming that a single effective theory with fixed
parameters remains valid throughout a wide range of energy or curvature
scales. (Further problems have been pointed out in
\cite{DiracPoly,TransCommAs}.)  In addition, the constructions presented in
\cite{Transfig} have helped to reveal the full scope of the covariance problem
in models of loop quantum gravity. The model of \cite{Transfig} quantizes not
only the homogeneous black-hole interior, but also applies homogeneous
minisuperspace models to the exterior, using timelike slices which are
homogeneous in a static space-time region. However, the same symmetry must be
compatible with spherically symmetric, inhomogeneous, spacelike slices. For
the vacuum models considered in \cite{Transfig}, dilaton gravity theories
present a powerful tool to analyze the possible dynamical equations consistent
with holonomy-modified minisuperspace models on timelike slices.

In a more cosmological context, we may view the spherically symmetric exterior
of a non-rotating black hole as a static version of Lemaitre--Tolman--Bondi
(LTB) models
\begin{equation} \label{LTB}
 {\rm d}s^2= -M^2{\rm d}T^2+ S^2{\rm d}X^2+ R^2({\rm
   d}\vartheta^2+\sin^2\vartheta {\rm d}\varphi^2)
\end{equation}
with $T$-independent $M$, $S$ and $R$. As proposed in \cite{Transfig}, one can
turn the staticity condition, together with spherical symmetry, into a
homogeneity condition on timelike slices, $X={\rm const}$. Redefining the
metric components according to 
\begin{equation} \label{Mapping}
 |p_c|=R^2\quad,\quad p_b= MR \quad,\quad N=S\,.
\end{equation}
and renaming the coordinates $T=x$ and $X=t$, we obtain a homogeneous model
\begin{equation} \label{LRS}
  {\rm d}s^2= -\frac{p_b^2}{|p_c|}{\rm d}x^2+N^2{\rm d}t^2+ |p_c| ({\rm
   d}\vartheta^2+\sin^2\vartheta {\rm d}\varphi^2)
\end{equation}
of timelike slices, thus the unusual signs of ${\rm d}x^2$ and ${\rm d}t^2$.
(The variables $p_b$ and $p_c$ correspond to densitized-triad components of
the homogeneous model; see \cite{BHInt}.)  Equations of motion generated by a
(modified) minisuperspace Hamiltonian constraint then determine how
extrinsic-curvature components $b$ and $c$ are related to time derivatives of
$p_b$ and $p_c$, as well as $N$.

Because a covariant, slicing-independent theory would have a single
formulation with the same degrees of freedom for homogeneous timelike and
inhomogeneous but static, spherically symmetric slices in the same region, the
minisuperspace model shows what kind of degrees of freedom should be
present. In particular, there should be fields only for a metric or a triad
(but no scalar in the vacuum model), and they should be local because they are
not accompanied by any additional degrees of freedom in a holonomy-modified
minisuperspace model. General results from $1+1$-dimensional dilaton gravity
\cite{Strobl} show that possible covariant and local modifications of
spherically symmetric or LTB dynamics are much more restricted than
modifications of homogeneous models, which in loop-quantum-cosmology
descriptions of (\ref{LRS}) have been applied rather liberally.  In
particular, covariance allows only the freedom of choosing a dilaton potential
in the action or Hamiltonian --- a function only of the coefficient $R$ but
independent of $S$ and $M$ in (\ref{LTB}).

As shown by an explicit transformation of equations of motion from modified
homogeneous models to the coefficients defined in (\ref{LTB}), however, the
modifications implied by holonomies in minisuperspace models {\em cannot} be
written in this way \cite{TransComm}.  In particular, because the
identification (\ref{Mapping}) relates $p_b$ and $N$ to $M$ and $S$, any
holonomy modification, which by definition depends on $b$ or $c$ or both in a
non-linear way, implies corrections in the spherically symmetric theory that
depend on $M$ or $S$. This result is in conflict with the condition that the
only available choice in a covariant $1+1$-dimensional dilaton model is given
by the $R$-dependent dilaton potential. Therefore, the assumption that
holonomy modifications are compatible with slicing independence is ruled out
by contradiction.

The specific construction not only invalidates the black-hole models attempted
in \cite{Transfig}, it also highlights deep problems implied by holonomy
modifications: Even if one is interested only in cosmological models but not
in a timelike homogeneous slicing as in \cite{Transfig}, a covariant set of
modifications should be implementable in a consistent way in all
cases in which multiple slicings are possible, respecting the required
symmetries. The observations of \cite{TransComm} then rule out holonomy
modifications as ingredients of slicing-independent space-time theories. They
constitute a no-go theorem for the possibility of covariant holonomy
modifications.

A possible solution to this problem is to implement holonomy modifications in
an anomaly-free way which does not break any gauge transformations but may
deform the classical structure of hypersurface deformations given in
\cite{DiracHamGR,Regained}. Consistent deformations are possible in
spherically symmetric models with holonomy modifications
\cite{JR,LTBII,HigherSpatial,SphSymmOp,ActionHigher,DefGenBH,DefGenThesis},
but they imply a non-classical space-time structure which is related to
slicing independence only in some cases, and after field redefinitions
\cite{Normal,EffLine}. The latter feature not only resolves the contradiction
between holonomy modifications and covariance pointed out in \cite{TransComm},
it also shows why singularities can be resolved in loop quantum cosmology even
for matter obeying the usual energy conditions: Not only the dynamics but also
space-time structure become non-classical as a consequence of holonomy
modifications, unhinging the mathematical foundation of singularity theorems.
But at the same time, deformations of space-time structure complicate any
analysis of quantum space-time. In particular, they invalidate existing
attempts to derive parameters of an effective theory of loop quantum cosmology
from deparameterized models of full loop quantum cosmology, which implicitly
assume that holonomy modifications are compatible with slicing independence;
see also \cite{CQGRCov}.
 
\section{What's left?}

During the early stages of developments in a new research field, a certain
laxness in rigor is often accepted in order to make some progress in spite of
mounting obstacles. Preliminary observations then indicate how promising the
new field may be, and where detailed investigations should be initiated in
order to buttress the field's foundations. In this spirit, the Ashtekar school
has led to useful results in quantum cosmology.

However, with the benefit of hindsight, it turned out that none of the
specific assumptions made explicitly or implicitly by the Ashtekar school
can be put on firm ground. Instead, as indicated by the account of the
foundational developments in \cite{Status}, the Ashtekar school took a wrong
turn at a very early stage, and then found it impossible to correct its
course. Crucial mistakes thereby were enshrined in its very foundations. These
mistakes are not just inaccuracies in early models that might be amended by
better approximations. Rather, they violate basic principles of a general
physical nature and can be corrected only by eliminating the key ingredients
that defined the Ashtekar school in the first place.

The large number of quotations from \cite{Status} given here has demonstrated
the unfortunate prevalence of mistaken assumptions and claims. Let us finally
collect the main erroneous statements made just in the Discussion section of
\cite{Status}. As always in this critique, emphases in quoted passages are
from \cite{Status}.
\begin{itemize}
\item ``As we saw in sections II -- IV, although this brilliant vision [of
  Wheeler's] did not materialize in the WDW theory, it {\em is} realized in
  all the cosmological models that have been studied in detail in LQC. However
  the mechanism is much deeper than just the `finite width of the wave
  packet': the key lies in the quantum effects of geometry that descend from
  full LQG to the cosmological settings. These effects produce an unforeseen
  repulsive force. Away from the Planck regime the force is completely
  negligible. But it rises {\em very} quickly as curvature approaches the
  Planck scale, overwhelms the enormous gravitational attraction and causes
  the quantum bounce.'' This statement overemphasizes the role of quantum
  geometry, while it ignores the fact that fluctuation effects explain much of
  the volume and density bounds obtained in loop quantum cosmology. Potential
  singularity resolution in loop quantum cosmology is therefore not dissimilar
  from what has been found in certain Wheeler--DeWitt-type quantizations; see
  for instance \cite{QCPerfectBohm,NonSingBohmQC,SingBI,SingLTB}.  The mistake
  is repeated in ``{\em In LQC the repulsive force has its origin in quantum
    geometry rather than quantum matter and it always overwhelms the classical
    gravitational attraction.}'' which also overstates the prevalence of
  bounces, as non-bouncing solutions are possible in general loop quantum
  cosmology \cite{NonBouncing}.
\item ``To obtain good behavior in both the ultraviolet {\em and} the infrared
  requires a great deal of care and sufficient control on rather subtle
  conceptual and mathematical issues.'' erroneously assumes that a single
  effective theory must be used through a wide energy or curvature range.
\item In ``Finally, it is pleasing to see that even in models that are {\em
    not} exactly soluble, states that are semi-classical at a late initial
  time continue to remain sharply peaked throughout the low curvature
  domain. [\ldots] Initially this is surprising because of one's experience
  with the spread of wave functions in non-relativistic quantum mechanics.''
  no surprise is warranted because in this regime one is dealing with a
  macroscopic object. Conversely, ``The third notable feature is the powerful
  role of effective equations discussed in section V. As is not uncommon in
  physics, their domain of validity is much larger than one might have naively
  expected from the assumptions that go into their derivation. Specifically,
  in all models in which detailed simulations of {\em quantum} evolution have
  been carried out, wave functions which resemble coherent states at late
  times follow the dynamical trajectories given by effective equations even in
  the deep Planck regime.'' should have raised a severe warning. This
  statement hides the unmentioned (but wrong) assumption that macroscopic
  averaging regions may be used even ``in the deep Planck regime.''
\item The claim that effective equations ``arise from a (first order)
  covariant action'' is incorrect because the proposed action fails in vacuum
  models. Similarly, more advanced recent versions which include a scalar
  field \cite{LimCurvLQC,HigherDerivLQC} fail to describe anisotropic or
  inhomogeneous modes in congruence with loop quantum cosmology
  \cite{MimeticLQC,CovModPert,MimeticLQCPert,DefSchwarzschild2}.
\item ``The very considerable research in the BKL conjecture in general
  relativity suggests that, as generic space-like singularities are
  approached, `terms containing time derivatives in the dynamical equations
  dominate over those containing spatial derivatives' and dynamics of fields
  at any fixed spatial point is better and better described by the homogeneous
  Bianchi models. Therefore, to handle the Planck regime to an adequate
  approximation, it may well suffice to treat just the homogeneous modes using
  LQG and regard inhomogeneitys as small deviations propagating on the
  resulting homogeneous LQC {\em quantum} geometries.'' gives a correct
  qualitative description of the BKL scenario but then misapplies it by
  referring to entire homogeneous models rather than microscopic homogeneous
  regions. 
\item ``Returning to the more restricted setting of cosmology, it seems fair
  to say that LQC provides a coherent and conceptually complete paradigm that
  is free of the difficulties associated with the big-bang and
  big-crunch. Therefore, the field is now sufficiently mature to address
  observational issues.'' is premature, given the serious problems of the
  approach reviewed in \cite{Status}. In particular, quantization ambiguities,
  effective theory, and covariance must be under control for reliable
  observational predictions.
\end{itemize}

To summarize, defining assumptions made by the Ashtekar school have violated
basic concepts of effective field theory, which led it to overestimate the
robustness of their models and to downplay quantization ambiguities. It
oversimplified the approach to high curvature, leading to unreliable
statements about the quantum nature of the big bang. Even if a valid effective
theory could be derived from loop quantum gravity, it would not be expected to
confirm results by the Ashtekar school because an effective theory rarely
makes use directly of fundamental parameters, as assumed by the Ashtekar
school, and crucially includes running parameters. Morever, models developed
by the Ashtekar school cannot be covariant, as shown by a no-go theorem based
on the discussion from Section~\ref{s:Cov}. Violating such an important
consistency condition is not an approximation that may be improved by
including further corrections. Models that violate essential symmetries rather
produce spurious solutions which invalidate any analysis; see for instance
\cite{GaugeInvTransPlanck} in a different context. In loop quantum cosmology,
this outcome is shown by the possibility of signature change at high density
in non-classical space-time structures compatible with holonomy modifications
\cite{Action,SigChange,ScalarHolInv}. The transition through high density is
then no longer deterministic \cite{SigImpl,Loss}, a result which cannot be
obtained from deterministic bounce claims by adding small corrections.

We have reached a rather sobering end. It may be possible that some readers of
this critique do not agree with all the conclusions drawn here. But it should
have become clear that the present status of loop quantum cosmology can
accomodate a wide variety of interpretations, not just the overly optimistic
view espoused by the Ashtekar school \cite{Status}.  At present, it is not
possible to derive a sufficiently complete effective theory from full loop
quantum gravity, and even if this were possible, it is not clear whether loop
quantum gravity itself is covariant and consistent. Nevertheless, as sketched
in the present contribution, one can test the validity of proposed models
based on general requirements on effective theories, and independently perform
tests of covariance. Given presently available models, the precise behavior at
Planckian densities remains undetermined, and even qualitative features are
too ambiguous and uncertain to select a specific scenario. However, careful
statements about the approach to high curvature may be made, indicating what
implications the first deviations from classical behavior of gravity or
space-time might imply.

Bringing attention to the possibility of different outcomes in light of
ambiguities and uncertainties may at some point indicate a certain universal
behavior, which would be missed if one insists on a presupposed outcome. Such
questions may be more modest than the grand claims presented in \cite{Status},
but they are well worth pursuing.

\section*{Acknowledgements}

This work was supported in part by NSF grant PHY-1912168.


\end{document}